\documentclass[fleqn,10pt]{wlscirep}
\usepackage[utf8]{inputenc}
\usepackage[T1]{fontenc}
\usepackage{physics}
\usepackage{amsmath}
\usepackage{mathtools}
\DeclareMathOperator{\taninv}{tan^{-1}}
\title{Implementation of a Hamming-Distance-Like Genomic Quantum Classifier Using Inner Products on IBMQX4 and IBMQX16}

\author[1,a,*]{Kunal Kathuria}
\author[2]{Aakrosh Ratan}
\author[1]{Michael McConnell}
\author[1,b,*]{Stefan Bekiranov}

\affil[1]{Department of Biochemistry and Molecular Genetics, University of Virginia, Charlottesville, VA, USA}
\affil[2]{Center for Public Health Genomics, University of Virginia, Charlottesville, VA, USA}
\affil[*]{corresponding author}
\affil[a]{kk7t@virginia.edu}
\affil[b]{sb3de@virginia.edu}

\begin{abstract}
 Motivated by the problem of classifying individuals with a disease versus controls using functional genomic attributes as input, we encode the input as a string of 1s (presence) or 0s (absence) of the genomic attribute across the genome. Blocks of physical regions in the subdivided genome serve as the feature dimensions, which takes full advantage of remaining in the computational basis of a quantum computer. Given that a natural distance between two binary strings is the Hamming distance and that this distance shares properties with an inner product between qubits, we developed two Hamming-distance-like classifiers which apply two different kinds of inner products ("active" and "symmetric") to directly and efficiently classify a test input into either of two training classes. To account for multiple disease and control samples, each train class can be composed of an arbitrary number of bit strings (i.e., number of samples) that can be compressed into one input string per class. Thus, our circuits require the same number of qubits regardless of the number of training samples. These algorithms, which implement a training bisection decision plane, were simulated and implemented on IBMQX4 and IBMQX16. The latter allowed for encoding of $64$ training features across the genome for $2$ (disease and control) training classes.
\end{abstract}
\begin{document}

\flushbottom
\maketitle
%
%
\thispagestyle{empty}

\section*{Introduction}

Quantum computing algorithms have been developed that show great promise of making potentially significant improvements upon existing classical equivalents, particularly in the area of machine learning. Exponential speedups have been shown for implementing least squares fitting \cite{Biam9}, quantum Boltzmann machines \cite{Biam22,Biam61}, quantum principal components analysis \cite{Biam11} and quantum support vector machines \cite{Biam13} on a quantum computer, over their classical counterparts \cite{Biamonte2017}. Quadratic speedups have been demonstrated for Bayesian inference \cite{Biam106,Biam107}, online perceptron \cite{Biam108}, classical Boltzmann machines \cite{Biam20} and quantum reinforcement learning \cite{Biam30, Biamonte2017}. However, these speedups presume a low-error-rate, universal, quantum computer with hundreds to thousands of qubits. In addition, the speedup of a subset of these algorithms (e.g. quantum support vector machines) requires quantum RAM which would enable a quantum coherent mapping of a classical vector into a quantum state \cite{Lloyd2014}, but this does not currently exist. Recent progress has been made in exploiting a relatively natural connection between kernel-based classification and quantum computing \cite{Schuld2017, Hav2019, schuld2019}. One set of approaches maps a large feature space to a quantum state space in order to compute a kernel function on a quantum computer which is then used to optimally classify input data using, for example, a support vector machine (SVM) \cite{Hav2019, schuld2019}. In another set of approaches, variational quantum circuits are used to directly classify the data in the large quantum feature space on a quantum computer, similar to the approach employed by an SVM \cite{schuld2019, Hav2019}. An even more direct and simple implementation of kernel-based classification on quantum computers exploit quantum interference of the training and test vectors which when properly prepared as quantum states execute distance-based classification upon measurement \cite{Schuld2017, SchuldProc}. An advantage of these kernel-based approaches is that they can and have been implemented on existing quantum computers. However, in the case of the quantum interference circuit \cite{Schuld2017} the distance measure was Euclidean-based and only two-training vectors were used to build the classifier. Motivated by the problem of classifying individuals with a disease (e.g., Alzheimer's disease) given single cell neuronal genomic copy number variation (CNV) data \cite{vanden2016, McConnell2013,mcConnell2017,Chronister2019}, we developed a set of quantum classifier circuits which exploit a biologically relevant encoding of the training and test vectors that allow us to take full advantage of the computational basis of the quantum computer. Specifically, the training and test vectors encode the presence and absence of a CNV in a given genomic window with a 1 and 0, respectively, in non-overlapping windows across the human genome (i.e. they can be represented as binary strings). These genomic windows, which are nothing but ordered physical subdivisions of the genome, act as our feature dimensions (more precisely, the CNVs in respective genomic regions constitute the different features). The classification is efficient as it is based directly on the highest inner product between the test and training vector group of each class. This is possible because the class label qubit (which is eventually measured) is directly entangled with the training vectors, rendering the usual ancilla qubit unnecessary. Furthermore, we take advantage of many of the shared properties between a natural distance measure between binary strings, the Hamming distance, and the inner product between vectors composed of many binary strings to arrive at the simplest implementation of a near-term quantum disease-control classifier using genomic input data. The classifier is broad in its scope and can be applied to any situation amenable to using inner products in a feature space composed of divisions of physical space, time etc. or of any correspondingly comparable attributes as discussed in the following sections. \\

Our circuits are employed to execute binary (disease/normal) classification of genomic samples on 5-qubit (IBMQX4) and 14-qubit (IBMQX16) architectures using two different inner-product metrics. Genomics is an exciting rapidly advancing field that is much in need of effective machine-learning solutions to keep up with fast-growing technological advancements in sequencing, the large amount of data generated in each sequence and the vast array of different relevant data types. Development of these and other near-term quantum algorithms will enable quantum computation to solve demanding, data-driven problems in genomics as we approach the development of powerful, low-error-rate, universal quantum computers.\\

\section*{Results and Discussion}

\noindent
If one is interested in quantifying how well two arbitrary binary strings (of equal length $n$) match with each other, the most natural way would be to calculate their Hamming distance. The Hamming distance is simply the sum of the positional mismatches of the two bit strings. Thus, the Hamming distance of identical strings is $0$ and that of two strings that are binary complements is $n$. \\

\noindent
The Hamming distance has broad utility in solving classification problems that can use binary or binarized inputs representing a more complex event. Our classifier is similar in spirit to the Hamming distance as it is based on the inner product of bit strings. In fact, it is exactly equivalent to a Hamming-distance-based classifier provided some caveats to the input data (see Methods for proof). As a simple warm-up example without any technical implementation details, we examine how our inner-product-based classifier can be applied to a schedule-matching problem. There are certain time blocks in the day where individuals are either available or not available, and the goal is to determine which individual (or set of individuals) a given "test" individual's schedule matches best (so they can find the most common time among all individual pairs to work together on a new issue, for example). Each time block is represented uniquely by a "block state" and a coefficient/prefactor attached to each such state represents a person's availability during that time of day. For example, if we had $8$ time blocks, we would represent them with $3$ qubits, where the block state $\ket{000}$ would represent the first of these $8$ blocks and the block state $\ket{111}$ would represent the last block. Thus the time blocks constitute the feature space of the problem. In the simplest case, the state coefficients would be binary themselves ($1$ indicating the person's availability and $0$ indicating non-availability). \\

\noindent
The prefactored block states for each individual are summed to yield that individual's total state vector. An inner product between two individuals' state vectors would then yield a score quantifying how well the two individual schedules agree. A class is defined generally as a collection of state vectors (henceforth referred to as "training vectors") satisfying a unique grouping property or pattern. The training vectors of each class are summed to yield its \emph{class vector}. The classifier simultaneously executes an inner product between the test input and each class vector, which is equal to the sum of the inner products between the test vector and the class's training vectors. Based on measurement probabilities, the classifier then classifies the test input into the training class whose class vector yields the highest inner product with the test. In general the class vector is allowed to be a vector sum of many individuals' state vectors that belong to the same class (and is prepared based on a classically precomputed sum for the actual circuit). Let us suppose for this example that there were $3$ schedules in all, $2$ training schedules represented by vectors $A$ and $B$ each playing the role of a class vector for its respective class, and $1$ test input represented by vector $T$. Suppose that there are $3$ time blocks and $A$ has an opening only in the first time block, $B$ only in the second and $T$ in both the second and the third. Thus $\mathbf{A}$ = (1,0,0), $\mathbf{B}$=(0,1,0) and $\mathbf{T}=(0,1,1)$. We immediately see that $\mathbf{T}\cdot\mathbf{B} > \mathbf{T}\cdot\mathbf{A}$ and therefore the test vector is classified into the class of class vector $\bf{B}$. This classification is exactly implemented by our inner product classifier. This problem is of course just one example of a broad range of possible applications for this classifier. \\

\noindent
The specific motivation for the development of our classifier was a DNA copy-number-variation(CNV)-based disease classification problem in genomics. The human genome (which is a collection of chromosomes) can be subdivided into smaller regional blocks in chromosomal coordinate space. This is done effectively in the reference genome, which is a consensus of a relatively small number of individual's genomes and is represented as one large sequence of "A," "T," "C" and/or "G" bases strung together in a line. Each coordinate block of this reference genome is marked for the presence or absence of a CNV as found in a sequenced individual sample genome (any given sample genome itself is not as well-studied as the reference and does not have a robust coordinate representation by itself, and hence is "mapped" to the reference coordinate space. For technical details, see for example ref. \cite{Trapnell2009}.). A CNV in a given regional block indicates a deviation from the number of times (referred to as "copy number") the genomic string/sequence of that region in the reference genome is expected to occur in the sample genome. The default expected copy-number of any region is $2$ (there are two copies of each chromosome inherited by every individual from both parents). Copy-number variations are associated with a variety of phenotypes and can be strongly correlated with various diseases \cite{Lafrate2004}. We looked at CNVs in neuronal cell samples from healthy individuals as well as those affected by Alzheimer's Disease (AD) and developed this classifier as an attempt to classify genomes as healthy or containing AD. This is in fact a very natural specific application for our generic classifier in the regional/spatial domain. \\

\begin{figure}
    \centering
    \includegraphics[width=\linewidth]{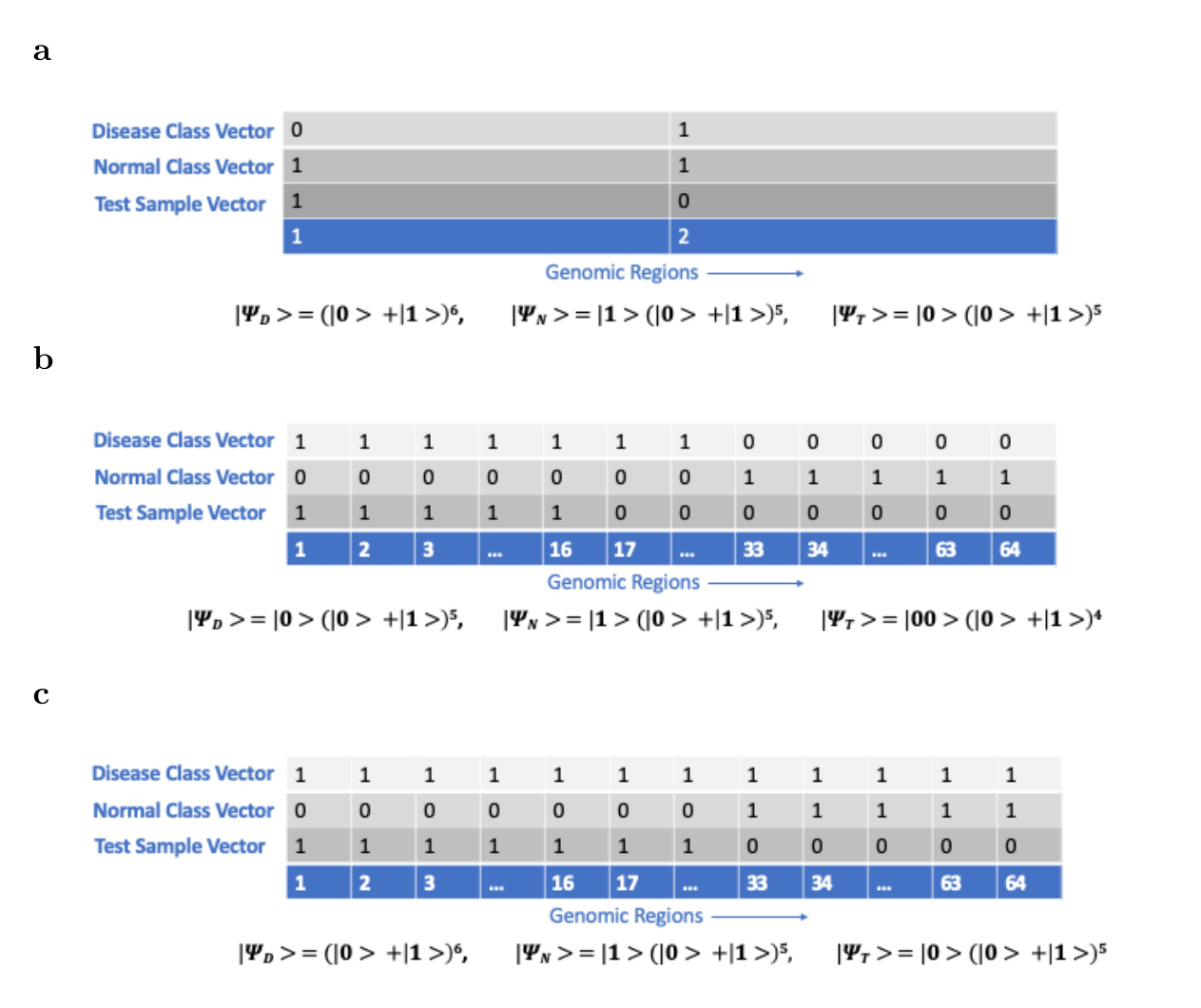}
    \caption{Examples of feature-space vectors of genomic samples containing different CNV patterns. Each column in the genome represents a feature dimension or physical genomic region. "1" indicates the presence of a CNV in a given region, and "0" indicates its absence. \textbf{a} shows a case where there are two genomic regions (setup of 5-qubit Example Problem 1), and \textbf{b} and \textbf{c} show 64 genomic regions (setup of 14-qubit Example Problem 1 and 2 respectively). Ellipsis points in a dimensional label imply that the vector has the same CNV value for all the unindicated dimensions. Unnormalized wavefunctions for disease and normal class vectors, as well as for the test vector, are written with subscripts "D," "N" and "T" respectively. The unnormalized wavefunctions simply show summed block states in the computational basis pertaining to the respective vector qubits only (not class, swapper etc.). They are all shown in the AIP framework for simplicity, though the situation depicted in (c) is solved in the manuscript using the SIP framework. }
    \label{fig:CNVSetup}
\end{figure}

\noindent
Thus the genome is divided into regional blocks (see Fig. \ref{fig:CNVSetup}) and each block state represents exactly one such region in order. Similar to above, if we subdivided the reference into $64$ genomic regions, we would represent them with $6$ qubits where a block state of $\ket{000000}$ would represent the $1^\textrm{st}$ region and the block state $\ket{111111}$ would represent the $64^\textrm{th}$ region. Thus, we will identify our ordered $n$-qubit computational "basis vectors" (or block states) with basis vectors in feature space in order to encode training and test vectors (referred to as "sample vectors") in the usual fashion \cite{Schuld2017}. If there is a copy-number-variation in the $i^{th}$ genomic region of a sample, the sample vector has a projection of $1$ onto the $i^{th}$ feature dimension, and if a CNV is absent there, its projection onto the same dimension is $0$. As an example, if we had a sample with $4$ genomic regions such that the first three contained a CNV and the last one did not, the sample vector in the feature basis would be $S=(1,1,1,0)$.\\
 
\subsection*{The Classification Metrics}

\noindent
We employ two classification metrics: the active inner product (AIP) and the symmetric inner product (SIP). First we will list some common features shared by the two metrics. Both metrics have the special merit of being able to handle an arbitrary number of training samples/inputs in the summed class-vector form. This is because both are linear: the sum of the inner products between the test vector and each training vector is the same as the inner product between the test vector and the class vector (see Methods for proof). This means that in one mathematical operation arbitrarily many inner products can be calculated between the test and training vectors and summed together for each class. This is the implementational advantage of our classifier over a Hamming-distance- based classifier, as the Hamming distance is not a linear measure for multiple bit strings in the sense above and would not allow for simultaneous calculation of arbitrarily many mutual distances. Further, for both the inner products, the sample vectors are easily represented in the computational basis using only unitary (or zero-valued) prefactors. As a note on efficiency, the swap-test applied to $n$-qubit states can simultaneously evaluate the inner product of two $2^n$-dimensional vectors in the feature basis, which is the formulation we use, without the need to store each vector component separately. This has the potential to become much faster than the classical inner product, which in the worst case will compute $2^n$ vector-component products and add them together. Finally, in our framework, there is no theoretical limitation as to the number of dimensions/regions encoded in feature space or the kind of binary-level/CNV-level training data that can be encoded. The difference between the two classification metrics lies in the initial state preparation and not in the inner product implementation itself. \\

\noindent
The classes of the training data represent in our case disease and normal samples. If the data is separable in some sense, the training vectors for the disease class will have a different regional block pattern for CNV presence as compared to those of the normal class. Given a test sample and sufficient separable training data, both the classifiers would classify the test sample into the appropriate category, making disease diagnosis of unknown random samples thus possible. As a technical note, the CNV-level data at this initial stage of the classifier does not allow for distinctions between genomic deletions and duplications. \\

\subsubsection*{Metric 1: Active Inner Product}

\noindent
The active inner product (AIP) is defined as the \emph{total} number of times the same region in the test and training vectors for a given class contains a CNV. It is simply the dot product between the sample and class vector in the feature basis. For example, if one of our class vectors was $C=(5,4,3,2)$ and the test vector was $T=(0,1,1,0)$ in the feature basis, the AIP between them would be $7$ (for simplicity, we are ignoring normalization). The test sample is then classified into the class yielding the higher inner product with this test vector. The sample vectors are encoded in the computational basis just as they are in the feature basis, without any change in their respective components. Thus, for each sample vector, all block states that correspond to genomic regions containing a CNV are assigned a coefficient of $1$ whereas regions not containing a CNV are assigned a coefficient of $0$. These block states are summed in the usual way to produce the state vector for each sample in the computational basis. For example, if there were a single sample divided into four genomic regions and only the first $3$ had a CNV each, its normalized state vector $\ket{\psi}$ would be: \\

\begin{equation}
    \ket{\psi} = \frac{1}{\sqrt{3}} \left ( \ket{00} + \ket{01} + \ket{10} \right ) \label{AIPsample}
\end{equation}

\noindent
where the state $\ket{11}$ is not present due to a coefficient of $0$. And if there were a class with two such identical samples, its class vector under the AIP framework in the computational basis would read (using $\psi$ as a generic placeholder again): \\

\begin{equation}
    \ket{\psi} = \frac{1}{\sqrt{12}} \left ( 2 \ket{00} + 2 \ket{01} + 2 \ket{10} \right ) \label{AIPClass}
\end{equation}

\noindent
The inner product is calculated between the test vector and the class vector (the vector sum of the training vectors in each class) simultaneously for the two training classes classes by overlapping their states via the swap test \cite{swaptest}. The test sample is classified now into its rightful class (implementation details shown shortly). We note that AIP gives preference to "$1$-matches" over "$0$-matches." \\

Though both AIP and SIP (described next) are well suited for our genomic problem depending upon the context, AIP is the naturally applicable metric for the schedule-matching problem. In fact, it happens to be even better suited than the Hamming distance for that problems of that nature. This is because in such cases one is interested in how well people's availabilities match, which the AIP calculates by summing the total number of  $1$-matches without any regard for the non-availabilities or $0$-matches (unlike Hamming distance). For our genomic problem, though the Hamming distance is the natural classifier of choice, AIP may be better suited for samples or diseases where the existence of a CNV in the same region in different samples is considered more significant than the absence of a CNV.  However, even if this CNV-bias were absent, the data that we deal with in such contexts is typically in a form such that AIP turns out to be equivalent to the Hamming distance as a classification metric as previously mentioned (see Methods). \\

\subsubsection*{Metric 2: Symmetric Inner Product}

\noindent
The symmetric inner product (SIP) is defined as the total number of times the same region in the test and training vectors for a given class "match" in terms of CNV presence, minus the number of times they do not match. "Matching" refers to two vectors both having a CNV or not having a CNV in the same region. In the SIP framework, the sample vectors are represented a bit differently in the computational basis as compared to the feature basis. For each sample vector, all block states that correspond to genomic regions containing a CNV are assigned a coefficient of $1$ whereas regions not containing a CNV are assigned a coefficient of $-1$. For example, the sample vector in \eqref{AIPsample} would be represented by: \\

\begin{equation}
    \ket{\psi} = \frac{1}{2} \left ( \ket{00} + \ket{01} + \ket{10} - \ket{11} \right ) \label{SIPsample}
\end{equation}

The rest of the routine proceeds identically to the AIP routine. The reason for the coefficient $-1$ in the SIP is that it introduces a penalty for unlike CNV events. Two non-CNV regions or two CNV regions will both contribute $+1$ to the inner product, but one CNV and one non-CNV region will contribute $-1$ to the inner product to account for mismatches in regional CNV events. SIP is exactly equivalent to Hamming distance as a classification measure (see Methods) and is thus very well suited to CNV-based genome classification (and many other similar problems). \\

\subsection*{The Inner-Product Decision Plane}

\begin{figure}
    \centering
    \includegraphics[width=\linewidth]{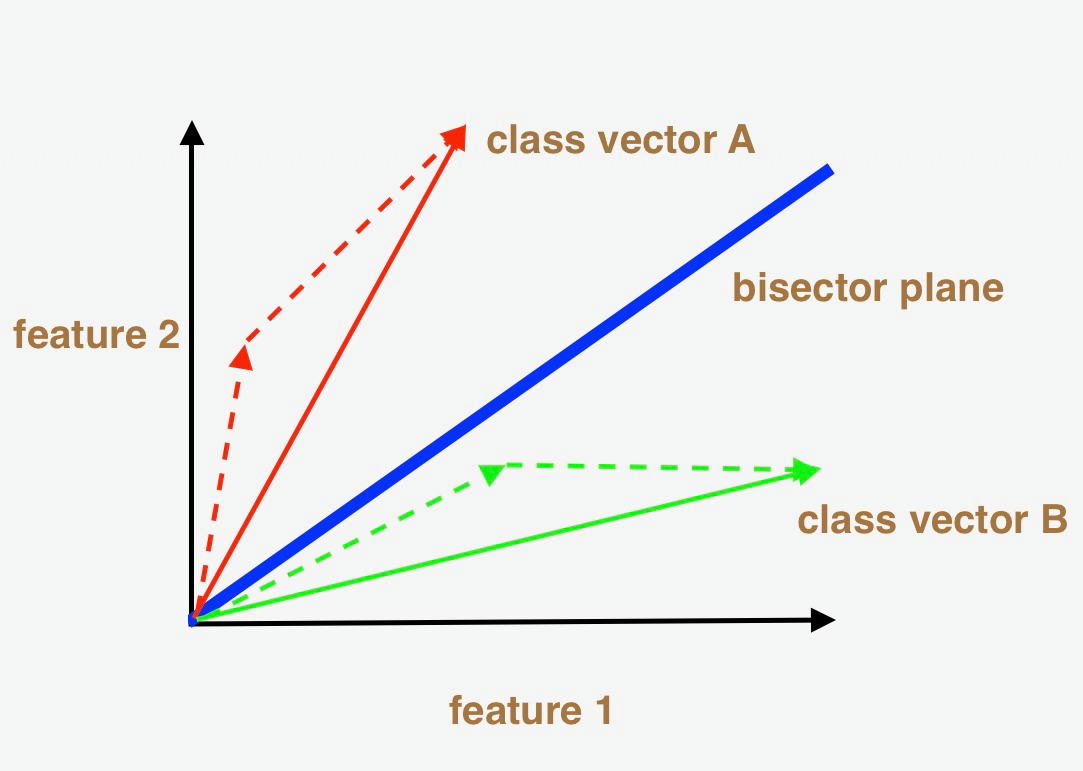}
    \caption{Inner Product Decision Plane in $2$ feature dimensions. There are two classes represented in red and green respectively. The class vector for each class is the sum of the class's respective training vectors, shown with dotted lines. The decision plane, shown in blue, bisects the two class vectors. The test vector will of course be classified according to the side of the decision plane in which it is located. For simplicity, only positive feature coordinates are shown.}
    \label{fig:DecPlane}
\end{figure}

\noindent
The decision plane for both the active and the symmetric inner product is the same. As shown in Fig. \ref{fig:DecPlane} for 2 feature dimensions, the decision plane is the bisector plane of the two class vectors, as the test vector is classified along with the class vector with which it yields a higher dot product. This of course means that its projection onto that vector is larger than its projection onto the other class vector. For a higher number of feature dimensions, the decision plane is the bisecting hyperplane that is orthogonal to the plane in which the class vectors lie. One advantage of this formulation is that one effectively compares the test vector with arbitrarily many training vectors for each class in one mathematical operation, as these training vectors are summed together into one final class vector for each class. This is possible due to the linearity of the inner product metrics we employ (more in following section). As an aside, for multiple class vectors, the decision boundary is not a simple hyperplane and is elaborated upon in the Supplementary Notes. \\

\subsection*{The Inner Product Circuit}

\noindent
The state manipulation routines and circuits can be formulated as generic functions of data parameters (see Methods), but we utilize circuit optimizations as suitable to the form of input data in any given problem. In terms of future vision, when the equivalent stage of "quantum hardware assembly" arrives, optimizations pertaining to specific kinds of data operations would hopefully be effectively applied in some shape in the design of the quantum processor. We will now look at the different general components/stages of the inner product circuit. In the state equations of this section, we note that the order of the states on the R.H.S. corresponds to the physical order of the qubits that the states occupy on the real hardware (this will be relevant for upcoming swap operations as part of the inner product calculation). In addition, the subscript \emph{following} a state denotes the qubits' functional role (as opposed to physical qubit position above) and is written in order of the qubits that compose the state vector. The class vector qubit(s) is (are) labelled by the subscript "d," the class index qubit by "m," the test vector qubit(s) by "t" and the swapper qubit by "s." For example, the two-qubit state $\ket{00}_{md}$ denotes that the "left" qubit in the state serves to encode the class index and the "right" one serves to encode the class vector. Further, when $\ket{\psi}$ is used with a subscript, it represents the state of the circuit when \emph{only} the qubits referred to by the subscript labels are considered (e.g. $\ket{\psi}_{md}$). When $\ket{\psi}$ is used with a superscript ($0$,$f$ or $c$), it represents the temporal state of all combined circuit elements ("initial","final" or "current" in the given context respectively). \\ 

\noindent
We now write the generic initial state for all inner-product circuits. The initial state is given by: \\

\begin{equation}
\ket{\psi^0} = \ket{s}_s\ket{t}_t\sum_{i=1}^M \ket{i}_m \ket{d_i}_d \frac{1}{\sqrt{M}} \label{psi0}
\end{equation}

\noindent
where $\ket{s}$ is the state of the swapper qubit, $\ket{t}$ is the state of the test vector, $\ket{i}$ indexes the computational basis state representing the $i^{th}$ class, $\ket{d_i}$ is the \emph{normalized} class vector state for the $i^{th}$ class, $M$ is the total number of classes and $\frac{1}{\sqrt{M}}$ is an overall normalization constant stemming from the number of classes. The "s" etc. labels used both in the subscript and state label itself are not redundant as they serve different purposes and are retained for clarity. Now, $\ket{d_i}$ can be specifically written in the feature basis as $\ket{d_i} = \sum_{f=1}^F c_f^i \ket{f}$, where $f$ represents a feature dimension, $c_f^i$'s are the projections of the class vector onto the $f^{th}$ feature dimension, $F$ is the total number of feature dimensions and $\ket{f}$ is a placeholder for the computational basis-vector state for the $f^{th}$ feature dimension. Analogously, $\ket{t}=\sum_{f=1}^F c_f^t \ket{f}$. Of course,  $\log_2 F$ qubits are required in general to index $F$ feature dimensions. A key point in the circuit is that the usual role of the ancilla qubit (see for example \cite{Schuld2017}) is subsumed in the class index qubit (we only need $1$ qubit to index $2$ input classes), which is now directly entangled with the summed training vectors via the sum in \eqref{psi0}. \\

\noindent
All circuits proceed in the same general way through the following stages:
\begin{enumerate}
    \item The feature-space data is encoded in the circuit.
    \item The well-known swap test is applied to calculate the inner product between the test and entangled class vectors.
    \item The probabilities encoding the final inner product result are measured in the computational basis.
\end{enumerate}

As mentioned, the data-encoding may vary from circuit to circuit and will be specifically presented with the example problems. The Fredkin (controlled swap) gate is used in the swap test as presented in ref. \cite{Smolin1996}. The swapper qubit $s$ is acted upon by a Hadamard gate, used as a control qubit to swap $\ket{t}_t$ and $\sum_{i=1}^M\ket{d_i}_d$ via the Fredkin gate, and again acted upon by a Hadamard gate. This effectively calculates the inner product between the states contained in the qubits it swaps (in our case the test vector and the class vectors). For clarity, the exact inner product between the test and the entangled class vectors is given by: \\

\begin{align}
\braket{t}{\phi} = \sum_{i=1}^M \sum_{f=1}^F c_f^t c_f^i \ket{i}_m \frac{1}{\sqrt{M}} \label{IP}
\end{align}

\noindent
where $\ket{\phi}_{md} \equiv \sum_{i=1}^M \ket{i}_m \ket{d_i}_d \frac{1}{\sqrt{M}}$ and the overlap above is understood to be between the test state and the class vector state, i.e. "between" indices "t" and "d." We will soon see how this closely matches the monotonically equivalent form of the inner product the circuit calculates. \\

\noindent
The final state vector after the swap test reads (please note the ordering of states): \\

\begin{equation}
\ket{\psi^f} = \frac{1}{2}\ket{0}_s(\ket{t}_t \sum_{i=1}^M \ket{i}_m \ket{d_i}_d \cdot \frac{1}{\sqrt{M}} + \sum_{i=1}^M \ket{i}_m \ket{d_i}_d  \ket{t}_t \cdot \frac{1}{\sqrt{M}}) + \frac{1}{2}\ket{1}_s(\ket{t}_t \sum_{i=1}^M \ket{i}_m \ket{d_i}_d \cdot \frac{1}{\sqrt{M}} - \sum_{i=1}^M \ket{i}_m \ket{d_i}_d  \ket{t}_t \cdot \frac{1}{\sqrt{M}})
\end{equation}

\noindent
The interchange of the order of the test and class-training states in each term in parentheses signifies the swapping of the qubits holding these state vectors (see examples below for more details). Repeated measurements of the swapper and class index qubits are made to arrive at the measurement probability encoding the solution to our problem. To see how this is so, let us calculate some relevant theoretical measurement probabilities. Define $\rho_{\overline{s}k}$ as the probability of measuring the swapper qubit in state $\ket{s=\overline{s}}$ and the class index qubit in state $\ket{i=k}$. For reasons pertaining to better statistical fidelity observed when compared to using $\overline{s}=0$, we will focus on the measurement probabilities $\rho_{00}$ and $\rho_{11}$. For detailed analysis of error modes associated with IBM processors, see for example ref. \cite{Sisodia2017}. First, the projection operator projecting the above state onto states corresponding to fixed values of the swapper ($s=1$) and class index ($i=k$), the $\ket{s=1}_s \ket{i=k}_m$ basis, yields:\\

\begin{align}
P(\ket{s=1}_s,\ket{i=k}_m) \ket{\psi^f} &= \frac{1}{2 \sqrt{M}}\ket{1}_s \left ( \ket{t}_t \sum_{i=1}^M \delta_{ik} \ket{i}_m \ket{d_i}_d  - \sum_{i=1}^M \delta_{ik} \ket{i}_m \ket{d_i}_d \ket{t}_t \right ) \nonumber \\
&= \frac{1}{2 \sqrt{M}}\ket{1}_s \bigg( \ket{t}_t \ket{k}_m \ket{d_k}_d  - \ket{k}_m \ket{d_k}_d \ket{t}_t \bigg)
\end{align}

\noindent
where the Kronecker delta function $\delta_{ik}$ serves to pick out terms corresponding to class $k$. The squared norm of the above is \\

\begin{align}
\rho_{1k} &= \frac{1}{4 M}\bra{1}_s \bigg( (\bra{t}_t \bra{k}_m \bra{d_k}_d - \bra{k}_m.\bra{d_k}_d \bra{t}_t \bigg) \ket{1}_s \bigg( (\ket{t}_t \ket{k}_m \ket{d_k}_d - \ket{k}_m.\ket{d_k}_d \ket{t}_t \bigg) \nonumber \\
&= \frac{1}{2M} \bigg( 1 - |A_k|^2 \bigg) \label{Pr}
\end{align}

\noindent
where $A_k \equiv \braket{t}{d_k}$ is exactly equal to the inner product between the test vector and the class vector of class $k$. Thus, the directly measurable probability $\rho_{1k}$ is the negative of the inner product added to a constant, always a positive number (as is easily verified in \eqref{Pr}) and a monotonically decreasing function of the true inner product of normalized states. All the circuits thus end with a measurement of the probabilities $\rho_{10}$ and $\rho_{11}$ and our classifier choosing the class $k$ that yields the lower measured probability and thus the higher inner product with the test vector. \\

\noindent
In terms of data-encoding, though specific routines for encoding data will be employed in our circuits, the kind of data that can be entered in principle is not limited (up to normalization constraints mentioned in Methods). The general recipe for encoding arbitrary data involves solving a system of non-linear equations. However, circuits encoding non-trivial data may have extremely large gate depth in large part due to the limitations of CNOT connectivity in currently available quantum processors. Therefore, for relevance and clarity, we will present problems involving simplified data that can be input feasibly into IBM machines. These solutions themselves require non-trivial gate-depth at times as they need several swap operations. \\

\noindent
We now present our circuits with their respective example problems. We will first describe the 5-qubit circuit and then the 14-qubit version. For each, problems are solved using the simulator as well as the real processor using the AIP and/or SIP framework as applicable. \footnote{Problems that are solved by the simulator are also solvable by the real processors, which present additional architectural constraints, and we present such solution counterparts for most of the cases.} For each case, we will first introduce the generic circuit and state notation, and then present specific example problems. \\

\subsection*{Five-qubit Generic Circuit}
\noindent
AIP problems involving an artificial genome containing two regional blocks can be solved with the 5-qubit circuit. The SIP framework does not have much relevance or power here due to the small feature space. Only 1 qubit was used to encode the class vector ("d"), which spans the two genomic regions. The other 3 qubits were taken by the class index ("m"), test vector ("t"), which also resides in 2-dimensional genomic feature space, and the swapper qubit ("s"). \\

\noindent
As we have 1 qubit to index the class, $\ket{0}_m$ is the first computational basis state representing class $0$ and $\ket{1}_m$ the second representing class $1$. This class index qubit is first entangled with the training qubit using a "rotation routine" so as to separate the training vectors for each class. As an example, the state $\ket{\psi}_{md} = \ket{00}_{md} + \ket{11}_{md}$ represents the simple case that in class $0$ (disease class) there is a training vector with a CNV only in the first of two regions (block state $\ket{0}_d$) and in class $1$ (normal class) there is a CNV only in the second region given by block state $\ket{1}_d$. (using the qubit labelling convention described previously). \\

\subsubsection*{Example Problem 1: AIP on 2-Block Genome}
We will solve a very simple problem with the 5-qubit circuit now in the AIP framework. In this example the normal class vector (class $1$) has one training vector that contains a CNV in each of the 2 total genomic regions. The disease class vector (class $0$) has only 1 CNV, present in the second region. The test vector also has only 1 CNV, present in the first region (see Fig. \ref{fig:CNVSetup}(a)). Thus, the test sample should be classified as normal. \\

\begin{figure}
    \centering
    \includegraphics[width=\linewidth]{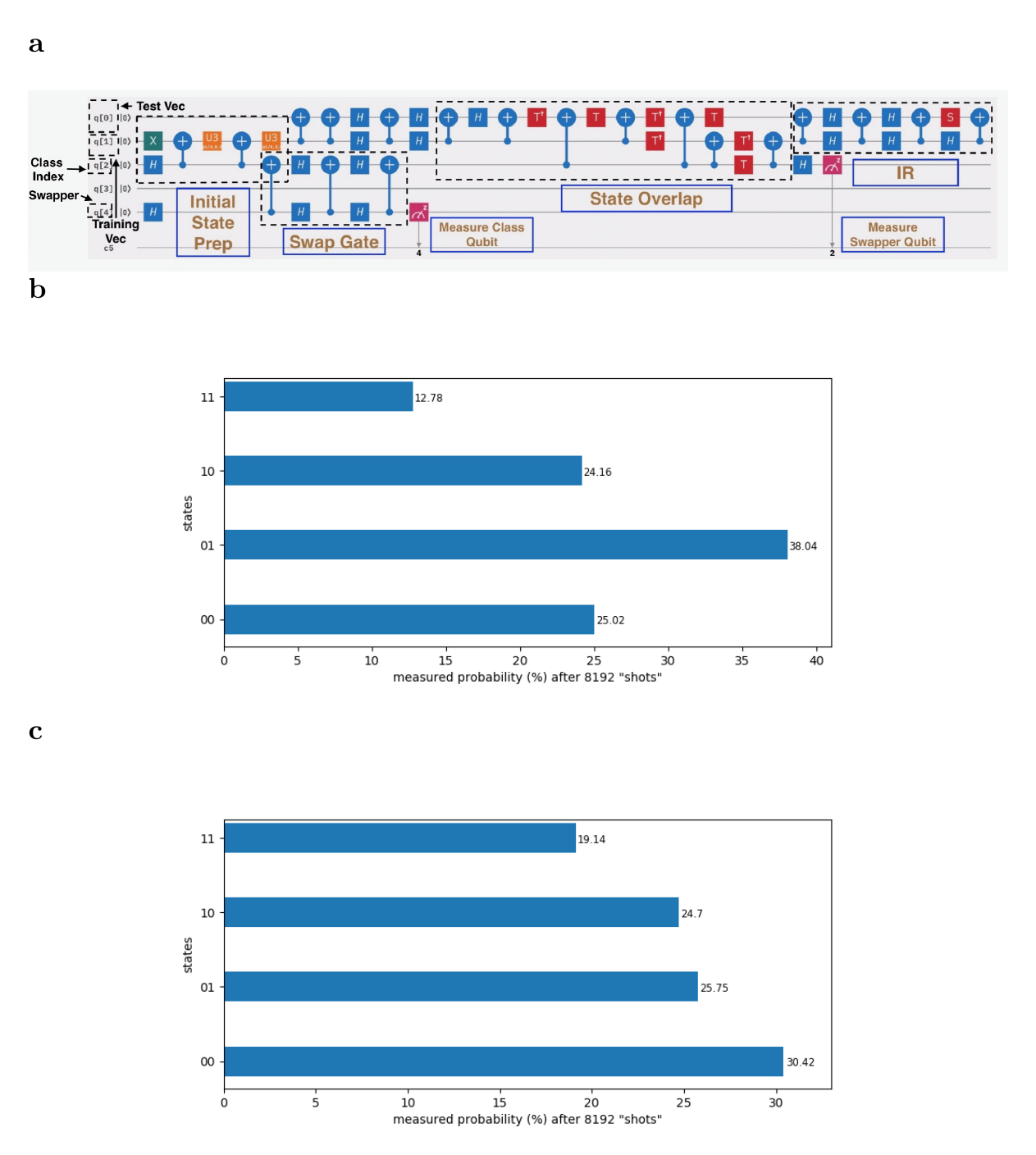}
    \caption{5-qubit Example Problem 1 as drawn on IBM Composer. \\ We show \textbf{(a)} the circuit, \textbf{(b)} measurement probabilities on simulator and \textbf{(c)} measurement probabilities on IBMQX4. In \textbf{(a)}, sequential circuit modules are boxed and labelled. The swap gate, which is a necessity for IBMQX4 CNOT-connectivity constraints, swaps the qubits for the class index and swapper qubits during execution; the "IR" box contains gates that do not alter the measurement probabilities as they are placed after measurement, but are phase-altering components from the Fredkin gate retained for completeness;"CI" stands for "class index". In (b) and (c), the measured values of only the swapper and class index qubit states are shown for clarity in that order.}
    \label{fig:5qAll}
\end{figure}

\noindent
All qubits start in the ground state $\ket{0}$. To encode this data, a specific rotation routine is defined similar to other works \cite{Schuld2017} which allows one to prepare a broad class of entangled two-qubit states involving the class and training qubits. In these states, the disease class vector always has one regional block that contains a CNV while the other does not, and the normal class vector contains CNVs in both regions in any desired mutual ratio, $R_c$. Concretely, $R_c$ is the ratio of the number of CNVs present in the normal class vector in the first region to the number present in the second region (in this case, $R_c, for no reason other than simplicity, was chosen to be 1$). Our routine is applied to the initial state $\ket{\psi^0}_{md} = \ket{00}_{md}$ to render it into the generic state $\ket{\psi^f}_{md}= \ket{01}_{md} + \sine{\theta}\ket{10}_{md} + \cos{\theta}\ket{11}_{md}$ after employing six 1 qubit gates (as shown in module labelled "Initial State Preparation" in Fig. \ref{fig:5qAll}(a)). This signifies that class $0$ has a training vector that lies wholly in the second genomic region whereas class $1$ has a training vector with $\sine{\theta}$ CNVs in the first region and $\cos{\theta}$ CNVs in the second region, up to a normalization scaling factor (see Methods). The "rotation angle" $\theta$ can thus simply be written as $\theta = \taninv{R_c}$. The absence of a CNV in the first region of the disease class vector implies that the block state $\ket{00}_{md}$ does not feature above. Setting $R_c=1$ yields the post-rotation-routine state for this circuit: $\ket{\psi^c}_{md}=\frac{1}{\sqrt{2}}\ket{01}_{md} + \frac{1}{2} (\ket{10}_{md} + \ket{11}_{md})$. \\

\noindent
At this point, the qubit states for the other qubits are: $\ket{t} = \ket{0}_t$ (corresponding to the desired test sample vector) and $\ket{s} = \ket{0}_s$, yielding the state: \\

\begin{align}
\ket{\psi^c} = \ket{0}_s \ket{0}_t \left(\frac{1}{\sqrt{2}}\ket{0}_m\ket{1}_d + \frac{1}{\sqrt{2}}\ket{1}_m\frac{1}{\sqrt{2}}(\ket{0}_d + \ket{1}_d) \right)
\end{align}

\noindent
The second stage of the circuit is the swap test now effected on the training and test qubits with the swapper qubit as control, which encapsulates the inner product between the class vector and test vector qubit states (see Fig. \ref{fig:5qAll}). This inner product value is given by the measurement probabilities of the state of the "s" and "m" bits as measured in the computational basis, which leads us to the last phase of the computation. $8192$ "shots" are taken, and $\rho_{10}^M$ and $\rho_{11}^M$ are measured and plotted on a histogram (the superscript "M" denotes a measured probability as opposed to a theoretical one). As a prior, the theoretical values for the inner product are set by \eqref{Pr}, yielding $\rho_{10} = \frac{1}{4}$ and $\rho_{11} = \frac{1}{8}$, showing that the test vector yields a higher inner product with class $1$ and should be classified as such. \\

\begin{table}[ht]
\centering
\begin{tabular}{|l|l|l|l|}
\hline
Problem & Theory & Simulation & Real \\
\hline
5-qubit Example 1 & $\frac{1}{2}$ & 0.53 & .77\\
\hline
14-qubit Example 1 & $2$ & 1.9 & .26 \\
\hline
14-qubit Example 2 & $0$ & 0.0 & n/a \\
\hline
\end{tabular}
\caption{\label{tab:ResComparison} The ratio $\frac{\rho_{11}}{\rho_{10}}$}
\end{table}

\noindent
This problem is solved first on the IBM simulator on a version of the circuit not limited by architectural constraints, then on the simulator using a circuit appropriate for the IBMQX4 architecture, and finally using the same circuit on the IBMQX4 processor itself. The latter circuit along with the histogram of measured probabilities for the simulation and the real execution are shown in Fig. \ref{fig:5qAll}. The unconstrained simulation circuit is shown in Supplementary Figure 2. The real and simulated circuits differ only in the order of the qubits used to encode the data, and in the swap operations required to satisfy the constraints of the real processor. In Table \ref{tab:ResComparison} we compare probability values from the theoretical calculation, "measurements" from the architecturally-constrained simulation and measurements from the real execution for the various (5-qubit and 14-qubit) circuit examples. The simulation Fig. \ref{fig:5qAll}(b) yields $\frac{\rho_{11}}{\rho_{10}} = \frac{12.78}{24.16} = 0.53$ while IBMQX4 in Fig. \ref{fig:5qAll}(c) gives $\frac{\rho_{11}}{\rho_{10}} = \frac{19.14}{24.7} = 0.77$. As expected, the simulation probability is almost identical with the theoretical one for this example and classifies the test sample correctly as normal. The real measured probability, though $45\%$ away from the expected value, classifies the sample correctly. \\

\subsection*{Fourteen-qubit Generic Circuit}
\noindent
With the 14-qubit circuit, we can solve problems involving a simple genome containing up to $2^6=64$ regional blocks (with some optimizations this can be extended if there are any CNV-absent regions -- see Zero-coefficient Exclusion in Methods). Using the same notation scheme as for the 5-qubit circuit, 1 qubit was used to encode the class index, 6 qubits were used for the class vectors (labelled individually by "d1,...,d6" in decreasing order of bit place-value, or simply by "d" collectively), 6 qubits for the test vector ("t1,...,t6" individually or "t" collectively) and 1 for the swapper qubit ("s"). \\

\subsubsection*{Example Problem 1: AIP on 64-Block Genome}

\noindent
For this problem a single CNV is present in each of the first $32$ regions ($1-32$) only in the class vector for class $0$, and each of the last $32$ regions ($33-64$) only in the class vector for class $1$. The test vector represents a test sample that contains a CNV in each of regions $1-16$ only (see Fig. \ref{fig:CNVSetup}(b)). As a practical matter, the class vectors of this (and the following) example would be composed of multiple sample vectors to have this particular form in our context. The test sample here should be classified into class $0$ as a disease sample.\\

\begin{figure}
    \centering
    \includegraphics[width=.9\linewidth]{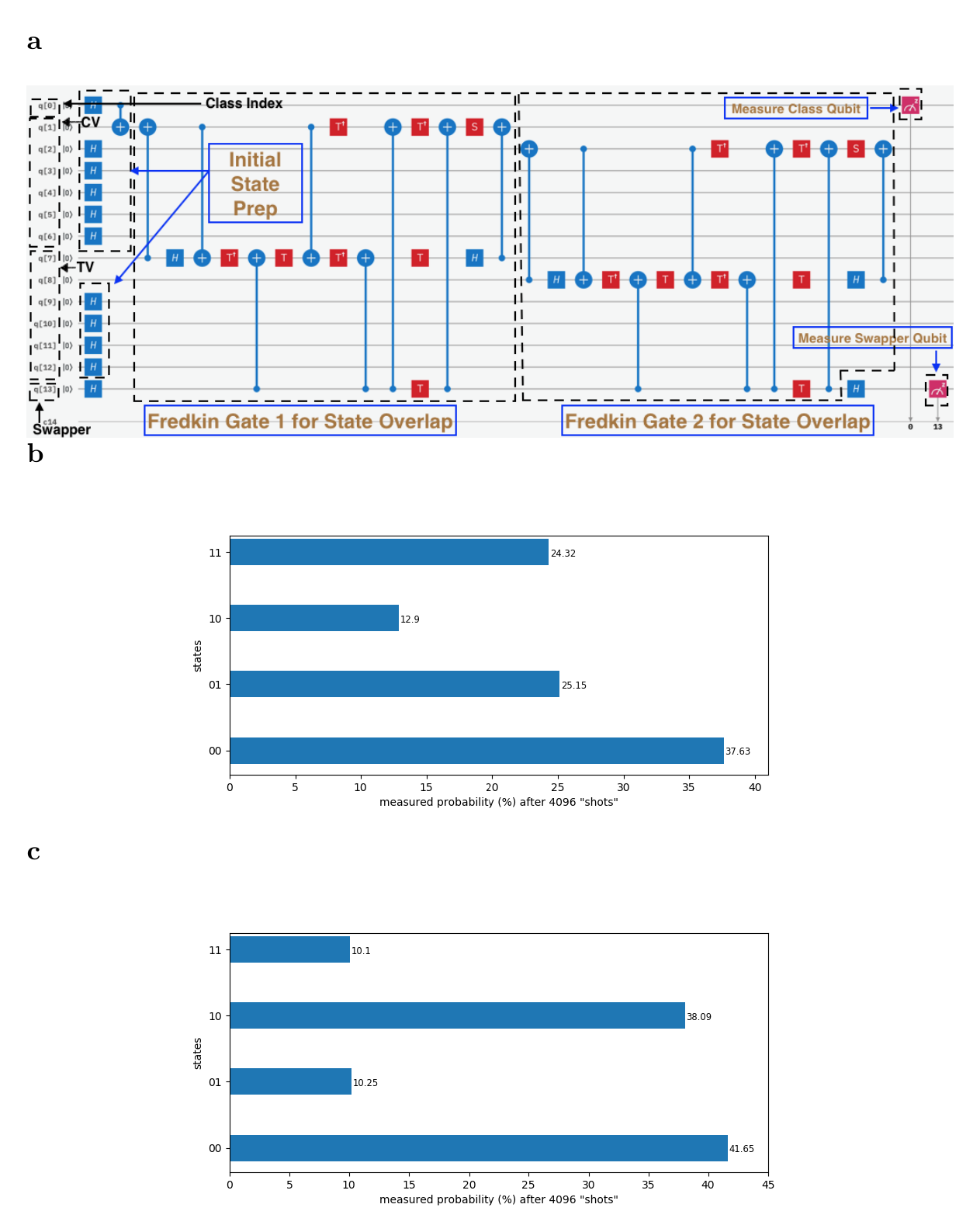}
    \caption{14-qubit Example Problem 1 as drawn on IBM Composer. \\ We show \textbf{(a)} the simulator circuit, \textbf{(b)} measurement probabilities on simulator and \textbf{(c)} measurement probabilities on IBMQX16. The circuit adapted to satisfy IBMQX16 connectivity constraints is too cumbersome to show here and is shown in Supplementary Figure 1. The measurement probabilities shown pertain the adapted version. "CV" refers to "class vector" and "TV" to "test vector." In the histograms, only the measured values of the swapper and class index qubit states are shown for clarity in that order.}
    \label{fig:14qAll1}
\end{figure}

\noindent
As previously, $\ket{0}_m$ is the first computational basis state representing class $0$ and $\ket{1}_m$ the second representing class $1$. As a note, in the simulated version of the problem, the first qubit ($q0$) is used for class index, $q1-q6$ for class vectors, $q7-q12$ for the test vector (both in decreasing order of place value of bits), and $q13$ for the swapper qubit. This changes in the real-processor version as shown in Supplementary Figure 1 for ease of required swap operations. The class index qubit is first entangled with the 6 training qubits as follows (see also Fig. \ref{fig:14qAll1}). A Hadamard gate is applied to $m$, followed by a CNOT gate with $m$ in control and $d1$ as target. This puts the first 7 qubits in the following superposition, effectively entangling the class index qubit with the 6 training qubits:\\

\begin{align}
    \ket{\psi}_{md} = \frac{1}{\sqrt{2}}\left( \ket{0}_m \ket{000000}_d + \ket{1}_m \ket{1000000}_d \right)
\end{align}

\noindent
To encode our data optimally, a Hadamard gate is applied to $d2-d6$ each and \emph{not} to $d1$. This puts $\ket{\psi}_{md}$ in the state: \\

\begin{align}
    \ket{\psi}_{md} = \left(\frac{1}{\sqrt{2}}\right)^6 \left( \ket{0}_m \ket{0}_{d1}(\ket{0}+\ket{1})^{\otimes 5}_{d2-d6} + \ket{1}_m \ket{1}_{d1} (\ket{0}+\ket{1})^{\otimes 5}_{d2-d6} \right)
\end{align}

\noindent
where the subscripted tensor product $(\ket{0}+\ket{1})^{\otimes 5}_{d2-d6} \equiv (\ket{0}+\ket{1})_{d2}(\ket{0}+\ket{1})_{d3}(\ket{0}+\ket{1})_{d4}(\ket{0}+\ket{1})_{d5}(\ket{0}+\ket{1})_{d6}$ etc. This state exactly represents the CNV configuration in the problem definition per the AIP framework (see Fig. \ref{fig:CNVSetup}). \\

\noindent
At this point, the qubit states for the other qubits are: $\ket{t} = \ket{000000}_t$ and $\ket{s} = \ket{0}_s$. Now, using the same technique as above to achieve the desired test vector, a Hadamard gate is applied to $t3-t6$ each, yielding the pre-swap-test state:\\

\begin{align}
\ket{\psi^c} = \left(\frac{1}{\sqrt{2}}\right)^{10} \ket{0}_s \ket{00}_{t1-t2}(\ket{0}+\ket{1})^{\otimes 4}_{t3-t6} \left( \ket{0}_m \ket{0}_{d1}(\ket{0}+\ket{1})^{\otimes 5}_{d2-d6} + \ket{1}_m \ket{1}_{d1} (\ket{0}+\ket{1})^{\otimes 5}_{d2-d6} \right)
\end{align}

\noindent
The swap test is executed as usual to evaluate the inner product via measurement probabilities of the states of the "s" and "m" bits in the computational basis. $8192$ "shots" are taken, and $\rho_{10}^M$ and $\rho_{11}^M$ are measured and plotted on histogram. In this case, the theoretical measurement probabilities from \eqref{Pr} are $\rho_{10} = \frac{1}{8}$ and $\rho_{11} = \frac{1}{4}$, showing that the test vector yields a higher inner product with class $1$ and should be classified as a disease sample. As before, the problem is first straightforwardly solved on the simulator, then on the simulator using an architecturally constrained circuit and finally on the IBMQX16 processor. The measured results along with the circuit are shown in Fig. \ref{fig:14qAll1}. Again, in Table \ref{tab:ResComparison} the theoretical probabilities, the architecturally simulated probabilities and the real measured probabilities can be seen. Both the simulated probabilities are very close to their respective theoretical values as expected. However, due to current limitations in hardware, measurements from the real processor yielded a value of $.26$ that did not show correlation with the theoretical value of $2$ and did not classify the sample correctly. \\

\noindent
As a last note, one can see in the circuit that the swap test was only applied to some of the test and class vector qubits and not to all. This is an optimization specific to this circuit as mentioned in "Inner Product Circuits" employing the principle of summing coefficient products of only like-valued qubits to calculate the inner product (please refer to Optimization Techniques in Methods). \\

\subsubsection*{Example Problem 2: SIP on 64-Block Genome}

\noindent
As shown in Fig. \ref{fig:CNVSetup}(c), for this problem a single CNV is present in each of the $64$ regions in the class vector for class $0$, and each of the last $32$ regions ($33-64$) only in the class vector for class $1$. The test vector represents a test sample that contains a CNV in each of regions $1-32$. Thus, the test sample should be classified into class $0$. As mentioned in Methods, one limitation of SIP is that one has to know a priori whether there are more total matches or total mismatches expected between the test vector and class vector in general, as only the square of the SIP, i.e., square of the difference between matches and mismatches, is actually measurable. This is no problem for the data we used to guide the development of our classification metrics \cite{vanden2016} (pioneering single-cell whole genome sequencing data for the context of our quantum classifier attempting to classify neuronal disease based on CNVs in samples, hereupon referred to as "contextual data"), where the number of matches always exceeds the number of mismatches (see SIP in Methods). This example is fabricated differently from the actual data to highlight certain features of the circuit \footnote{The data motivated the development of the classifier but example problems showing specific circuit functionality can certainly be set up with different-looking genomes} and here we will posit that the number of matches is not expected to exceed the number of mismatches, as shown in the genomic setup in Fig. \ref{fig:CNVSetup}(c). This means that the measured inner product between test and class-$1$ vector (stemming from an SIP of absolute value $64$, due to $0$ matches and $64$ mismatches) is expected to be higher than the inner product between test and class-$0$ vector (stemming from an SIP of $0$, due to $32$ matches and $32$ mismatches). \\

\begin{figure}
    \centering
    \includegraphics[width=\linewidth]{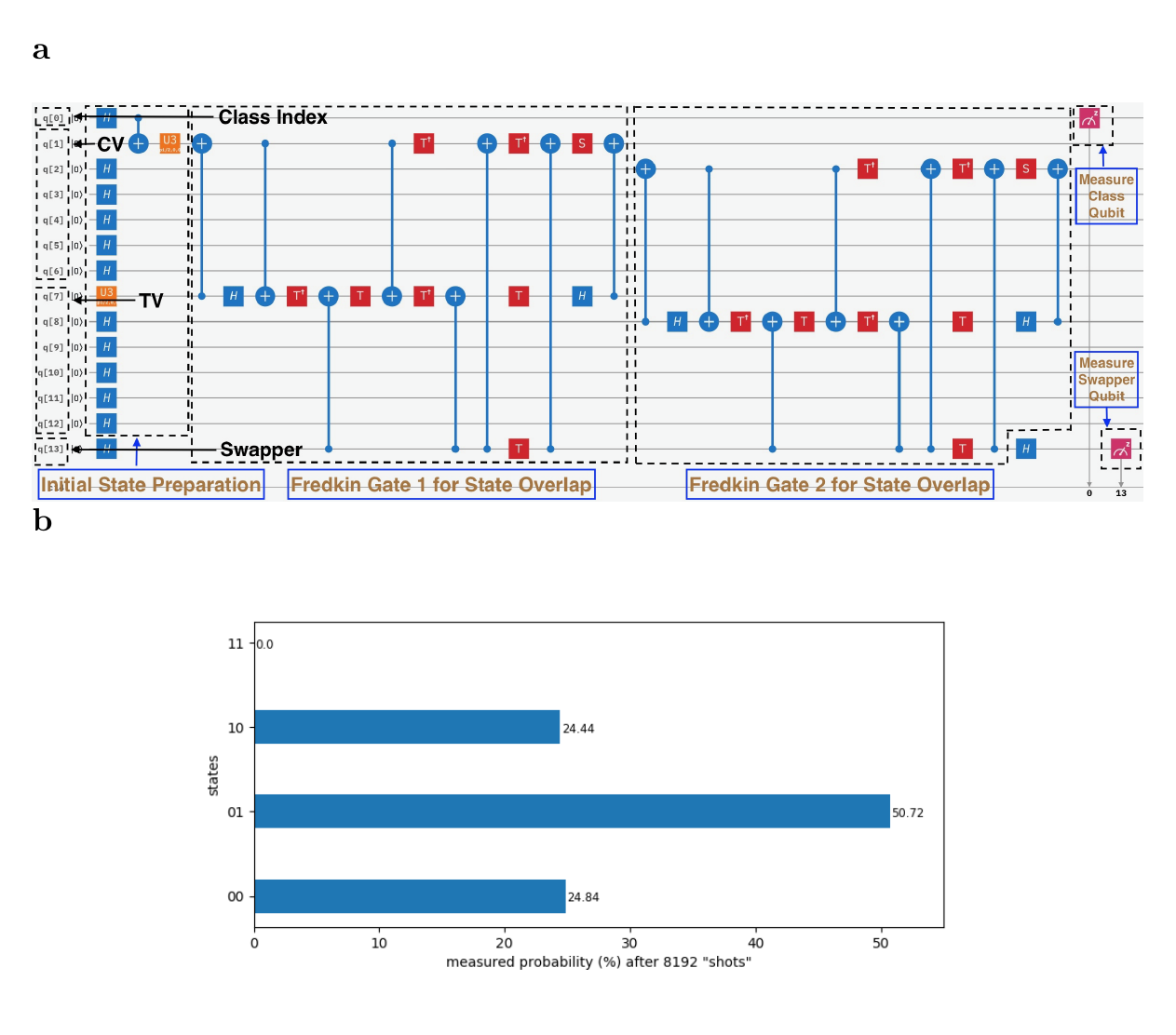}
    \caption{14-qubit Example Problem 2 as drawn on IBM Composer. \\ We show \textbf{(a)} the circuit and \textbf{(b)} measurement probabilities on simulator. "CV" refers to "class vector" and "TV" to "test vector." In the histograms, only the measured values of the swapper and class index qubit states are shown for clarity in that order.}
    \label{fig:14qAll2}
\end{figure}

In the same way as in the above example (see Fig. \ref{fig:14qAll2}), we create the state: \\

\begin{align}
    \ket{\psi}_{md} = \frac{1}{\sqrt{2}}\left( \ket{0}_m \ket{000000}_d + \ket{1}_m \ket{1000000}_d \right)
\end{align}

\noindent
and apply $R_y \left (\frac{\pi}{2}\right )$ to $d1$, and $R_y \left (-\frac{\pi}{2} \right )$ to $t1$, to arrive at the full state: \\

\begin{align}
\ket{\psi^c} = \left(\frac{1}{\sqrt{2}}\right)^{10} \ket{0}_s (\ket{0} - \ket{1})_{t1} (\ket{0}+\ket{1})^{\otimes 5}_{t2-t6} \left( \ket{0}_m (\ket{0}+\ket{1})^{\otimes 6}_{d1-d6} + \ket{1}_m (\ket{1}-\ket{0})_{d1} (\ket{0}+\ket{1})^{\otimes 5}_{d2-d6} \right). \\
\end{align}

The swap test was applied and the simulated probabilities measured as usual. We compare them with their theoretical counterparts in Table \ref{tab:ResComparison} and see that they are extremely close as expected. The test sample was classified into class $0$, which yielded the lower inner product, due to the precondition indicating fewer mismatches with class $0$. This example was not run on the real processor simply because, given current hardware limitations, it would add no value to our results. We have already demonstrated how a very similar design can be mounted on IBMQX16. \\

\section*{Methods}

\subsection*{Hamming Distance and Inner Product Equivalence}

All individual sample vectors (training or test) have binary-valued components (indicating in our context the presence or absence of a CNV in the regional dimensions of feature space). For what follows, we define the "bit-string-equivalent" (BSE) for a binary-valued sample vector such that the vector's $i^{th}$ component is simply treated as the value of the $i^{th}$ bit of its BSE. Then, the Hamming distance between two binary-valued sample vectors in feature space is naturally defined to be the Hamming distance between their respective BSEs. We will work out below the equivalence between the Hamming distance and the SIP/AIP of two sample vectors in the sense of their being equivalent classification measures. We will also show how SIP and AIP are linear in the sample vectors while the Hamming distance is not. \\

\subsubsection*{SIP}


We will first prove the linear behavior of SIP in terms of sample vectors and show how it is implemented as a veritable classifier in a quantum machine. Then, the proof of its equivalence with Hamming distance will follow. An $n$-dimensional class vector \emph{in feature space} takes the general form $C=\frac{1}{N_C}\sum_{f=1}^F c_f \vec{f}$, where  $\vec{f}$ is the indexed standard basis vector in $F$-dimensional feature space, $c_f$'s are un-normalized vector components or regional prefactors of $C$, and $N_C$ is an overall normalization constant. Suppose $C$ is composed of an arbitrary number of training vectors, which in analogous notation take the form $D=\frac{1}{N_D}\sum_{f=1}^F a_f^D \vec{f}$, where we note that $D$ serves as a training index and $a_f^D$ is a binary regional prefactor in feature space for the $D^{th}$ training vector. Suppose also that we have a test sample given by the vector $B=\frac{1}{N_B}\sum_{f=1}^F b_f \vec{f}$ with notation exactly analogous to $D$. The Hamming distance between $D$ and $B$ is now by definition: \\

\begin{equation}
H(B,D)=\sum_{f=1}^F(1-\delta_{a_f^D b_f}) \label{Hamming}
\end{equation}

\noindent
where $\delta_{a_f^D b_f}$ is the standard Kronecker delta function applied to $a_f^D$ and $b_f$. We are interested in minimizing the Hamming distance between sample vectors for optimal classification, which is equivalent to maximizing the quantity $-H$. \\

For clarity, we will focus on one training class only for this proof and generalize at the end. Let us first define $S_{11}(B,D)$ as the total number of times corresponding vector components/prefactors of $D$ and the test vector $B$ are both $1$, i.e. \\

\begin{equation}
S_{11}(B,D)= \sum_{f=1}^F \delta_{a_f^D 1} \delta_{b_f 1} \label{S_11}
\end{equation}

\noindent
Similarly, we define $S_{00}(B,D)$ as the total number of times corresponding components of $A$ and $B$ are both $0$, i.e. $S_{00}(B,D) = \sum_{f=1}^F \delta_{a_f^D 0} \delta_{b_f 0}$. We also define the "$1-0$ mismatches" in the self-evident way following above: $S_{01}(B,D) = \sum_{f=1}^F \delta_{a_f^D 0} \delta_{b_f 1}$ and $S_{10}(B,D) = \sum_{f=1}^F \delta_{a_f^D 1} \delta_{b_f 0}$. \\

Let us now define the quantity $S(B,D)$ as the total number of corresponding prefactor-matches minus the total number of prefactor-mismatches between the $D^{th}$ training vector and test vector $B$, i.e. $S(B,D)=S_{11}(B,D) + S_{00}(B,D) - S_{10}(B,D) - S_{01}(B,D)$. We note that SIP was defined as the natural extension of $S$, as the total number of prefactor matches in corresponding regions between the test vector and \emph{all} the training vectors in the class minus the total number of mismatches. Define $\sigma_{11}(B,C) = \sum_{D=1}^W S_{11}(B,D)$ etc. Thus, the SIP between test $B$ and class $C$ is written as: \\

\begin{align}
\sigma(B,C) &= \sigma_{11}(B,C) + \sigma_{00}(B,C) - \sigma_{10}(B,C) - \sigma_{01}(B,C) \label{sigmaSum} \\
&= \sum_{D=1}^W S(B,D) \label{SIP}
\end{align}

\noindent
where $W$ is the total number of training vectors in the class. In the computational basis, $\sigma$ is a natural linear extension of $S$ (i.e. the sum of the SIPs of multiple sample vectors is equivalent to the SIP of the sum of the vectors as far as successful classification is concerned.). Let us see how this condition of SIP \emph{linearity} is achieved and utilized in the computational basis, with state normalization, for the same above-given sample vectors. \\

\noindent
We will use different notation for some of the state vectors in this section compared to previous sections for ease of proof. In moving from feature basis to computational basis now, we will use corresponding Greek notation for state vectors where necessary. We recall that in the SIP framework, for the state vector of any training or test sample, the coefficient for the $f^{th}$ computational-basis vector can only be $1$ or $-1$ (pertaining to the case where there is a CNV present in the region and the case where there is not, respectively). So we start with defining the training vector (we will use $\ket{d}$ to denote the training vector and $\ket{w}$ to denote the class vector in this section to make the distinction): \\

\begin{equation}
    \ket{d} = \frac{1}{\eta_d} \left ( \sum_{f=1 | \alpha_f^d = 1}^F \alpha_f^d \ket{f} + \sum_{f=1 | \alpha_f^d = -1}^F \alpha_f^d \ket{f} \right ) \label{TrComp}
\end{equation}

\noindent
where $\ket{f}$'s are computational-basis vectors as before, $\alpha_f^d$ is the unnormalized $f^{th}$ component of $\ket{d}$ in the computational basis, $\eta_d$ is the overall normalization constant and $F$ again denotes the total number of feature/computational dimensions. The two summations in \eqref{TrComp} serve to pick out only terms where $\alpha_f^d=1$ and $\alpha_f^d=-1$ respectively.  Now the class vector is $\ket{w} = \frac{1}{\eta_w} \left ( \sum_{f=1 | \alpha_f^d = 1}^F  \sum_{d=1}^W \alpha_f^d \ket{f} + \sum_{f=1 | \alpha_f^d = -1}^F \sum_{d=1}^W \alpha_f^d \ket{f} \right ) $, where $\eta_w$ is the overall normalization constant and $W$ is again the total number of training vectors in the class. And finally, the test vector can be written as $\ket{t} = \frac{1}{\eta_t} \left ( \sum_{f=1 | \beta_f = 1}^F \beta_f \ket{f} + \sum_{f=1 | \beta_f = -1}^F \beta_f \ket{f} \right )$, in exact functional and notational analogy with $\ket{d}$ in \eqref{TrComp}. \\

\noindent
Given that the class vector is just a normalized sum over its training vectors, and that in any SIP in the computational basis each matching computational vector component will add $+1$ to the result and each mismatching component will add $-1$, $\braket{t}{d}$ is equal to the total number of regional CNV-matches minus the total number of mismatches between the test vector $\ket{t}$ and training vector $\ket{d}$, multiplied by their respective normalization constants. \\

\begin{align}
    \braket{t}{d} &= \frac{1}{\eta_d \eta_t} \left ( S_{11}(B,D) + S_{00}(B,D) - S_{01}(B,D) - S_{10}(B,D) \right ) \\
    &= \frac{1}{\eta_d \eta_t} S(B,D)
\end{align}

Now, since the class vector $\ket{w}$ is simply a linear sum of all its training vectors and the dot product (state overlap) is a linear operation: \\

\begin{align}
    \braket{t}{i} &= \frac{1}{\eta_w \eta_t} \sum_{D=1}^W S(B,D) \label{SIPLinearity} \\
    &= \frac{1}{\eta_w \eta_t} \sigma(B,C) \label{IP_TrClass_SIP}
\end{align}

\noindent
Note that the difference in normalization constant between the single training vector and the class vector is retained. As $W$ can be arbitrary, \eqref{SIPLinearity} proves the implementation of SIP linearity in the computational basis \footnote{The attentive reader will protest here as $\eta_w$ will change with addition of training vectors. This will not matter for classification if we keep the number of training vectors similar for both classes as seen next.}. We see that $\braket{t}{i}$ is of the form $\kappa \sigma(B,C)$, for some positive \emph{constant} $\kappa = \frac{1}{\eta_w \eta_t}$, and hence is a monotonic function of $\sigma(B,C)$. The fact that $\kappa$ is a constant is seen trivially for the case where there is only one training vector in each class, as $\eta_w = \frac{1}{\sqrt{n}}$ for both classes in the SIP framework. For the case of multiple training vectors per class, the form of the data determines the constancy of $\kappa$. The statistical nature of our neuronal AD/neurotypic contextual data is such that the total number of CNVs in each sample is similar across both disease and normal classes. In fact, the data reveals that the \emph{total} number of CNVs per genomic region has the same overall statistical distribution in each class \cite{vanden2016}. This gives rise to the normalization constant (which is a function of these regional coefficients) being the same for both classes in the computational basis. Classification in separable data is still possible due to a salient region where all training samples of a class show atypical activity or different regional arrangements giving rise to the same overall histogram of CNVs per region. As one general example, different floors of large multistory buildings may have different occupancy for different buildings, but the overall \emph{distribution} of floor occupancy could be very similar. At any rate, only the normalization constant needs to be similar for both classes. Any data satisfying this constraint is amenable to SIP classification. This implies that $\eta_w$, and thus $\kappa$, are very similar for both classes if the same (or similar) statistically significant number of training vectors is chosen for both, which is natural for our context. When $\braket{t}{i}$ is used as a classification metric, the higher of two SIPs will remain the higher one irrespective of the value of $\kappa$. This means that our classifier is valid independent of the value of the normalization constants. Thus, we have shown that our quantum machine successfully executes SIP-based classification and preserves linearity. \\

\noindent
There are ways to address the normalization constant issue and are expected to be presented in future work. As an example, one can encode a simplistic "scaling factor" (ratio of the raw normalization constant of one class vector to the other's) using the ratio of the coefficients of the ground and excited state of a qubit which would be multiplied and encoded into the inner-product at run-time. Of course, one can multiply this factor classically after the inner product computation, but that would not be a satisfying solution. Without this, if normalization constants for the classes are not similar, correct classification is still possible in a weaker sense. The normalization constant, being the square root of the sum of squares of basis-vector coefficients, is a slowly increasing function of any one coefficient. From the form of the inner product, one can see that if one genomic region in one training class had a particular prevalence of CNVs, classification of a similar test sample into that class would not suffer. In the contextual data, for example, a test sample typically has only $1$ CNV in the genome and is highly inclined towards being classified into the class displaying particularly more CNVs in the very same region where that CNV lies. \\

\noindent
Finally, we will now prove the equivalence of SIP to the Hamming distance (actually its negative, as mentioned) as a classification metric. In the language of SIP, \eqref{Hamming} can be rewritten by definition as $H(B,D)=S_{01}(B,D) + S_{10}(B,D)$. We will simply define the Hamming distance $\chi$ between a test vector and class vector to be the sum of the Hamming distances between the test vector and each training vector in the class, i.e. $\chi(B,C) = \sum_{D=1}^W H(B,D)$. As an aside here, we can immediately see how the Hamming distance is not a linear function of the sample vectors as $\chi$ is not a linear extension of $H$. As a trivial example, the two Hamming distances between $1$-d training vectors $(1),(0)$ and test vector $(1)$ add to $1$ whereas the Hamming distance between the test vector and the vector sum of the training vectors is $0$. Coming back to SIP-Hamming distance equivalence, $-\chi(B,C)$ can be rewritten as: \\

\begin{align}
    - \chi(B,C) &= - \sum_{D=1}^W S_{01}(B,D) + S_{10}(B,D) \\
    &= - \sum_{D=1}^W n - S_{11}(B,D) - S_{00}(B,D) \\
    &= \left ( \sum_{D=1}^W S_{11}(B,D) + S_{00}(B,D) \right ) - W n \label{negChi}
\end{align}

Now, \eqref{SIP} can be rewritten as: \\

\begin{align}
    \sigma(B,C) &= \sum_{D=1}^W S_{11}(B,D) + S_{00}(B,D) - (n - S_{00} (B,D) - S_{11}(B,D)) \\
    &= 2 \left ( \sum_{D=1}^W S_{11}(B,D) + S_{00}(B,D) \right ) - W n \label{sigmaRe}
\end{align}

\noindent
We see that \eqref{negChi} and \eqref{sigmaRe} are very similar. In each equation, if we fix the number of training vectors per class, the term $W n$ can be removed. Now, we see that 
$\sum_{D=1}^W S_{11}(B,D) + S_{00}(B,D)$ and $2 \sum_{D=1}^W S_{11}(B,D) + S_{00}(B,D)$ only differ by a factor of 2 and hence are classificationally equivalent. Thus we have proven that SIP and Hamming distance are classificationally equivalent measures. \\

\noindent
There is one main limitation of SIP to mention here. As measurement statistics are always represented by the square of the coefficients/probability amplitudes, $\braket{t}{i}$ in \eqref{IP_TrClass_SIP} will be measured indirectly via $|\braket{t}{i}|^2$. Thus, the missing information will be whether $\sigma(B,C)$ in \eqref{sigmaSum} is positive or negative, i.e. whether the total number of prefactor matches is higher or lower than the number of mismatches. Again, this is no issue for the contextual data since each sample has less than $2$ CNVs across all genomic regions \cite{vanden2016}, ensuring that any test sample will have many more matches (mostly "$0-0$ matches) than mismatches with each training class. For data where individual samples have decidedly fewer $1$'s than $0$'s, or vice-versa, the sign of $\sigma$ will be a priori known and this ambiguity will not be an issue. For clarity, we will summarize the cases where SIP-based classification in our circuits will not succeed: \\

\begin{enumerate}
    \item If the raw normalization constants of the two class vectors in feature space are not similar
    \item If the data differs by a significant overall scaling factor (see above or see Efficient Data-encoding Techniques)
    \item If it is not a priori known whether the number of prefactor matches will be greater or less than the number of prefactor mismatches
\end{enumerate}

\subsubsection*{AIP}

\noindent
AIP was defined as the total number of "$1-1$" matches between the test vector and all training vectors composing the class vector for a given class. Note that $\sigma_{11}(B,C)$ in \eqref{sigmaSum} is exactly the AIP, $A(B,C)$, for test $B$ and class $C$. As for SIP, we will again begin with showing how AIP is a bona fide classification metric and prove its linearity. In the AIP framework, we recall that in the state vector of any training or test sample the coefficient for the $f^{th}$ computational-basis vector can only be $1$ or $0$ (pertaining to the case where there is a CNV present in the region and the case where there is not, respectively). Following the same line of proof as for SIP and retaining the above feature basis definitions (and overall notation), we will now redefine state vectors in the computational basis. The training vector in the AIP framework reads: \\

\begin{align}
    \ket{d} &= \frac{1}{\eta_d} \left ( \sum_{f=1 | \alpha_f^d = 1}^F \alpha_f^d \ket{f} + \sum_{f=1 | \alpha_f^d = 0}^F \alpha_f^d \ket{f} \right ) \\
    &= \frac{1}{\eta_d} \sum_{f=1 | \alpha_f^d = 1}^F \alpha_f^d \ket{f}
\end{align}

\noindent 
The test vector can similarly be written as $\ket{t} = \frac{1}{\eta_t} \sum_{f=1 | \beta_f = 1}^F \beta_f \ket{f}$. Now, in analogy with SIP, \\

\begin{equation}
    \braket{t}{d} = \frac{1}{\eta_d \eta_t} S_{11}(B,D)
\end{equation}

and \\

\begin{align}
    \braket{t}{w} &= \frac{1}{\eta_w \eta_t} \sum_{D=1}^W S_{11}(B,D) \label{AIPLinearity} \\
    &= \frac{1}{\eta_w \eta_t} \sigma_{11}(B,C) \label{IP_TrClass_AIP}
\end{align}

\noindent
As $W$ can be arbitrary, \eqref{AIPLinearity} proves the implementation of AIP linearity in the computational basis. Following exactly the same argument as for SIP at this point for the same kind of data, we conclude that the above inner product leads to successful classification. Thus, we have shown that our quantum AIP machine classifies successfully and preserves linearity. \\

\noindent
We will now show the classificational equivalence of AIP to the negative Hamming distance with the caveat that its equivalence to Hamming distance is not crucial, even undesirable, for classification problems where $1-1$ matches are more significant than other kinds of prefactor matches. In such cases, like the schedule-matching problem shown in the previous section, AIP is the natural measure of choice. Proceeding with the demonstration of equivalence now, we know that the AIP between test sample $B$ and class vector of $C$, $\sigma_{11}(B,C) = \sum_{D=1}^W S_{11}(B,D)$. We would like to show as above that this is classificationally equivalent to $\sum_{D=1}^W S_{11}(B,D) + S_{00}(B,D) = \sigma_{11}(B,C) + \sigma_{00}(B,C)$. Clearly, this equivalence is valid when $\sigma_{00}$, the total number of "$0-0$" matches between the test and training vectors of the class in question, is either fixed for both classes somehow, or is a monotonically increasing function of $\sigma_{11}$ in general. The way these two conditions hold true would certainly vary from case to case, so we will simply sketch a few plausible scenarios here. The latter condition is true, for example, when the difference in the number of CNV and non-CNV regions in any given sample is similar to that in other samples (assuming again that the same number of training vectors is chosen for each class). It is also true in the case where CNVs are generally sparsely distributed for all samples. This is most certainly the situation with the contextual data, where any sample statistically has only a single CNV across the whole genome. For test and training samples with sparsely distributed CNVs, an increase in $\sigma_{11}$ is accompanied by an equal increase in $\sigma_{00}$ with a high likelihood. To see how this is the case, one can visualize a test sample vector with a CNV not "in line" (see Fig. \ref{fig:CNVSetup}) with regionally coinciding CNVs of a class's training vectors, and as soon as it is placed "in line" by swapping it with a non-CNV in another region in the sample, both the "$1-1$" and "$0-0$" matches will very likely simultaneously increase, as long as the likelihood of finding a CNV in the samples is low. Thus, we have shown that our AIP quantum machine is a valid classifier and that it is equivalent to Hamming-distance as a measure under certain conditions, and summarized specific cases satisfying those conditions. We will again enumerate the cases where AIP classification in our circuits will not succeed: \\ 

\begin{enumerate}
    \item If the raw normalization constants of the two class vectors in feature space are not similar
    \item If the data differs by a significant overall scaling factor (see Efficient Data-encoding Techniques)
    \item If the total number of "$0-0$" prefactor matches is neither fixed for the different classes nor is a monotonically increasing function of the "$1-1$" matches
\end{enumerate}

\subsection*{Efficient Data-encoding Techniques}

\noindent
One universal challenge for encoding feature-space data using the computational basis is that one can fix only the relative values of the vector components, as the state vector must be normalized in the computational basis. For example, the unnormalized vectors $A=(1,1), B=(2,2), C=(3,3)$ in two-dimensional feature space will all be encoded as $S=\frac{1}{\sqrt{2}}(1,1)$ in the computational basis and will not be distinguishable. As mentioned above, there are ways like encoding a scaling factor that future work would address. Fortunately, the kind of data that we are dealing with does not present this issue, again due to the fact that the raw normalization constant for the two classes is very similar if a similar number of training vectors is used. This normalization ambiguity is a current major limitation of our metrics when used for general classification problems without the scaling factor in place. Thus, the metrics are most applicable to data sets whose \emph{overall} scaling across the feature dimensions is similar enough so that the inner product values for the different classes would not be relatively affected. \\


\subsubsection*{Brute Force Approach}

\noindent
It is not in general trivial to encode arbitrary data values from feature space into $n$-qubit systems starting from the universal ground state via a series of available unitary gate operations. A non-trivial combination of ancilla-like entangling qubits and controlled rotations are typically required even for the case of two qubits (see "Initial State Preparation" in Fig. \ref{fig:5qAll}). Here, we will present a recipe for encoding arbitrary forms of data using entangled quit states, up to a possible quantifiable error, given normalization constraints. Typically, in our context, an $n$-qubit sample vector is encoded by the (binomial-series-like) state:

\begin{equation}
   \ket{\psi} =  (a_n\ket{0} + b_n \ket{1})...(a_1\ket{0} + b_1 \ket{1}) \label{series}
\end{equation}

\noindent
where the $a_i$'s ($i = {1,2,...n}$) are data-dependent coefficients with subscripts in decreasing order of place value. Even though this representation is without qubit entanglement, it has the potential to encode sample vectors otherwise represented by entangled states in the computational basis, up to some known error. As a reminder, the combined coefficient of the term $\ket{0}^n$ represents the prefactor (CNV-value) of the first physical (genomic) region, that of the term $\ket{0}^{n-1} \ket{1}$ represents the CNV-value of the second region etc. The recipe is as follows: \\

\begin{itemize}
\item If the sample vector can be represented by unentangled qubits in the computational basis, it is simply written in the form of \eqref{series} and all the coefficients $a_i$, $b_i$ in the solution $\ket{\psi}$ are readily identified. 
\item If not, first the sample vector is normalized in the feature basis using its overall normalization/scaling factor $\frac{1}{N}$. Then, expanding terms in \eqref{series} leads to $2^n$ non-linear equations in the coefficients $a_i$ and $b_i$ of the form $c_n c_{n-1}...c_1 = \frac{A}{N} \textrm{ etc.}$ where the $c_i$'s ($i = {1,2,...n}$) are placeholders for either $a_i$ or $b_i$, and $A$ refers to the relevant regional prefactor/CNV-value.
\item Next, the non-linear equations are numerically solved for the $a_i$ and $b_i$ in the closest possible form satisfying the constraint that each $i^{th}$ qubit state-term in $\ket{\psi}$ in \eqref{series} is individually normalized (clear below), which automatically leads to the solution $\ket{\psi}$ being normalized overall. Error minimization techniques can be used to solve these equations for the $a_i$'s and $b_i$'s with the state normalization constraint, possibly containing some error relative to the actual data. 
\item Finally, each $i^{th}$ qubit state  in \eqref{series} is rotated by $R_y(\theta)$ to achieve these minimum-error values of $a_i$ and $b_i$. 
\end{itemize}

\noindent
The last bullet reveals why the individual qubit states need to be normalized, so as to be achievable by a unitary gate operation on these qubits. Thus, modulo the scaling factor, this recipe can be used to encode a relatively large feature-data space up to some known possible error. There will necessarily be cases where a small error will not be achievable (e.g. cases where the natural representation of the sample vector in the computational basis is with entanglement and where our recipe leads to state coefficients that cannot be normalized without introducing a large error in the solution), rendering this recipe ineffective for such scenarios. \\

\subsubsection*{"Binomial Series" Approach}

\noindent
This is an optimization that can be utilized in both the AIP and SIP framework if the data has certain simplifying CNV patterns amenable to this approach. In fact, it was employed in both Example Problem 1 and 2 of the 14-qubit circuit. We first note that, in the AIP framework, the non-zero coefficients in \eqref{series} will decide which regional blocks are encoded as having CNVs and which ones as not. Therefore, the values of these coefficients can be chosen to encode specific themes or patterns in the data. We will first motivate this in the context of the AIP. \\

\noindent
Suppose, for example, that the data has all zero CNV entries in the first half (block) of all genomic regions (i.e. in 32 of 64 regions etc.). We immediately see that setting $a_n=0$ will achieve this condition. If the data has all zero entries in the second half of the first \emph{and} second half blocks of genomic regions, setting $b_{n-1}=0$ will achieve this. Setting $b_n,b_{n-1}=0$ will enable only states beginning with the term $\ket{00...}$ to be present, nullifying all CNV-values except those of the first half of the first half of all genomic regions and so on. Similarly if it is desired that the  CNV-values of the first half of all regions be exactly twice that of the second half, one would set $a_n=2$ and all the other coefficients to $1$ (prior to normalization) and so forth. Many such desired patterns can be creatively effected in data encoding from this general framework. For example, this approach was extended to the SIP framework of Example Problem 2 of the 14-qubit circuit by finding simple rotations to encode the zero coefficients as "$-1$." This "binomial" series approach combines the idea of using ordered bit strings/ block states to enumerate physical regions with the use of unentangled 1-qubit states juxtaposed to yield a series representation for all unentangled $n$-qubit states in the computational basis. This series representation is then minimalistically utilized for optimal encoding. \\

\subsection*{Optimization Techniques}

\subsubsection*{Swap Like-valued Bits Only in Inner Product Evaluation}

\noindent
When calculating an inner product of two state vectors, each composed of multiple qubits, one need not assess the contribution of corresponding qubits when their contribution is unity. This happens typically when corresponding qubits are like-valued. For example, if $\ket{A} = (a_1 \ket{0} + a_2 \ket{1})\ket{0}\ket{1}$ and $\ket{B} = (b_1 \ket{0} + b_2 \ket{1})\ket{0}\ket{1}$, 

\begin{align*}
    \braket{A}{B} &= a_1 b_1 \braket{0}{0} + a_1 b_2 \braket{0}{1} + a_2 b_1 \braket{1}{0} + a_2 b_2 \braket{1}{1} \\
    &= a_1 b_1 + a_2 b_2
\end{align*}

\noindent
Clearly, the second and third qubit values of $A$ and $B$ do not enter the actual evaluation of the inner product, as they are correspondingly identical and contribute a factor of $1$ to the result. Of course, if $a_1 = b_1$ and $a_2=b_2$, then the inner product is simply unity by default. For the kind of cases presented in Results, an efficient way to assess qubit equality is to apply a CNOT gate between two corresponding qubits, and then a NOT gate on the target (second) qubit, whose final value serves as a control to the swap test operation. A prior copy of the second qubit would be made via a CNOT gate having this qubit as control and a qubit containing $\ket{0}$ as the target, which would be the copy destination. In other words, the overlap between corresponding qubits is calculated only when they are different. The control qubit is $1$ only if the two qubits are different to begin with. The value of the second qubit is copied before the operation in order to preserve it. \\

\noindent
This is in line with the future vision for quantum computing mentioned in the previous section. There will no doubt be better and more efficient ways to determine qubit-state equality and to embed these optimizations within the quantum circuitry in the long term. We have provided a sketch here. We are currently unable to implement sophisticated circuits anyway due to architectural and other constraints, and in our work have given a proof of concept for future implementations of the inner-product classifier that will employ much more developed quantum hardware and circuitry. One could just as well insert swap gates for like-valued qubits (for computing their trivial overlaps) in addition to the unlike-valued case, in order to present "complete" circuits, but given the current state of the field we did not think it vital to the theme of this work. \\

\subsubsection*{Zero-coefficient Exclusion}

\noindent
This is a data-reformulation technique for AIP that saves dimensions in feature space and hence qubits in computational space. Suppose that some of the physical regions in the test vector have a prefactor of $0$, thus ensuring that these dimensions will have no contribution to the inner product. The feature vectors can now be mapped onto a new basis  so as to eliminate the non-contributing dimensions from the computational framework altogether. For example, say the original test vector resides in $4$-dimensional feature space as $\bf{x}=(1,0,1,0)$, requiring $2$ qubits to encode. Clearly, we only need to use dimensions $1$ and $3$ to calculate the AIP. Thus, we eliminate dimensions $2$ and $4$ from consideration and reformulate our basis such that the dimensional labels ${1,3} -> {1',2'}$, where the primed dimensions refer to the new basis. The test vector in the new basis now reads $\bf{x}'=(1,1)$, and is encoded in the usual way as $\frac{1}{\sqrt{2}}(1,1)$, requiring only $1$ qubit. The class vectors are also mapped to the new basis and the AIP is calculated in the usual manner. \\

\section*{Data availability}

The data that support the findings of this study are available from the corresponding author on reasonable request.

\section*{Acknowledgements}

We would like to thank Dr. Thomas Lehner and Dr. Geetha Senthil at the National Institute of Mental Health for stimulating our interest in this study, organizing a series of discussions which brought quantum computing, neuroscience, genomics and computational biology experts together to discuss the potential of quantum computing in solving computational challenges in neuroscience and supporting this work.  K.K. A.R. M.M. and S.B. were supported by NIH grant 3U01MH106882-04S1.  M.M. and A.R. were also supported by NIH grants 5U01MH106882-05 and P30CA044579 (to the UVA Cancer Center), respectively.

\section*{Author contributions statement}
 
 K.K. designed the classification metrics, quantum circuits and executed the study. S.B. conceived the study. K.K. and S.B. conceived inner product as a metric and performed the experiment. M.M. and A.R. set the biological context and motivated the biological problem.
 
 \section*{Competing interests}
 
 The authors declare that there are no competing interests.
 
\section*{Supplement}

\subsection*{Useful Techniques and Observations}

The following contains generally useful observations or techniques that the authors stumbled upon, but were not directly applied to our inner product circuits. \\

\subsubsection*{Quantum Compression Technique}
One can \emph{encode} an arbitrary large number $m$ using merely $2$ qubits as follows. Prepare a state $\ket{\psi} = a \ket{0} + b \ket{1}$ such that $a/b = k$, and another state $\ket{\phi} = c \ket{0} + d \ket{1}$ such that $c/d = l$, where $l \leq 1$ and $m = l 2^k$. Thus $m$ is a fraction, given by $l$, of the largest value that a binary string of $k$ bits can hold in the usual base-2 representation. Of course, $k$ could in principle be directly used to store the desired large value, but $k$ and $l$ are both used here for better precision. The success of quantum compression necessarily depends on the precision achievable in state preparation (i.e. in rotation angles). Classically, there is no way around storing the precise number itself in some form whereas here we can effectively compress it into one precisely prepared qubit. The measurement would constitute the decoding/decompression. \\

\subsubsection*{Calculating Hamming Distance with XOR-based Schemes}
We used the feature basis to encode our example data, but will point out here that the most seemingly natural way to calculate the Hamming distance in a quantum computer is to use bit-wise CNOT (or XOR) gates applied between bit strings that are encoded directly as a series of $\ket{0}$ and $\ket{1}$ states in the computational basis. For example, the Hamming distance between "$001$" and "$111$" is given by the bit string "$110$," which is of course output of $\ket{001} \oplus \ket{111}$. Measuring the coefficients of all 3-qubit basis states after this CNOT operation would reveal $\ket{110}$ as the final result. However, this way to calculate the Hamming distance is highly inefficient in general, as each bit occupies a qubit and each training vector has to be separately encoded. \\

\subsubsection*{Inner-product Decision Plane for Multiple Classes}
For multiple training classes, the decision boundary for the inner product classifiers is not a simple plane or hyperplane as it is for two classes. In order to construct the decision space, one would draw a bisector plane for each class, dividing the feature space into preferred subregions by class, and pick the class that is the universally preferred class in the region where the test vector lies. For example, on a plane, one could have $3$ class vectors for classes $1$, $2$ and $3$ respectively, yielding a total of $3$ bisecting planes or lines. These lines divide the feature space into "$R-1 > R-2$", "$R-2 > R-3$" etc. subregions for a total of six such overlapping subregions (see Supplementary Figure 3). Here, $R-1 > R-2$ means that class $1$ is preferred over class $2$ for that subregion. Each point in the feature space will be part of exactly two overlapping subregions where a particular class ($1$,$2$ or $3$) is preferred over the other two. So the point where test vector lies determines its classification without any ambiguity in this manner. \\

\subsection*{Supplementary Figures}

\setcounter{figure}{0}
\renewcommand{\figurename}{Supplementary Figure}

\begin{figure}[h]
    \centering
    \includegraphics[width = \linewidth]{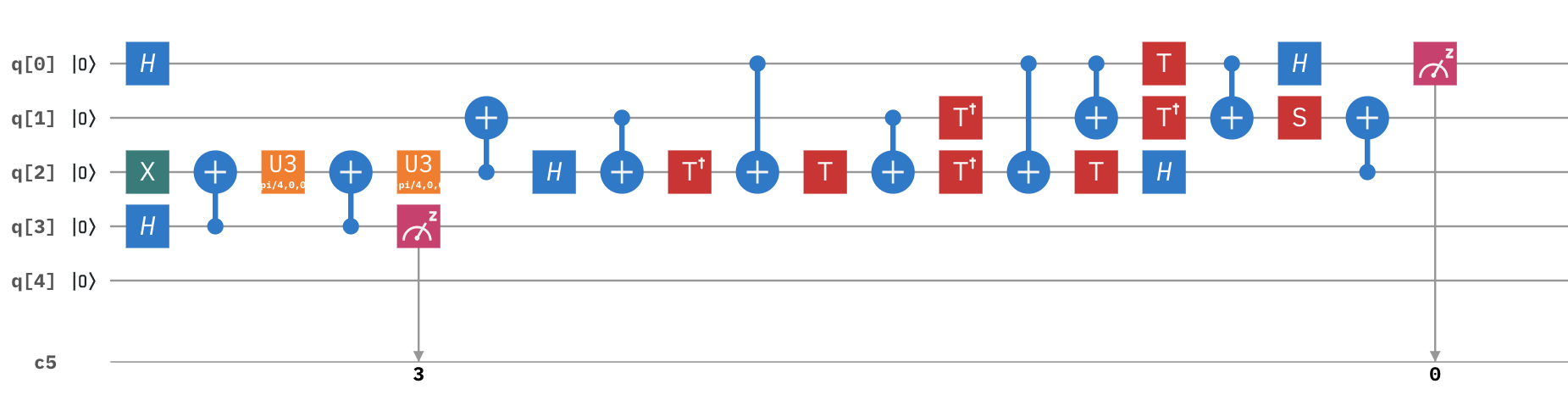}
    \caption{5-qubit Example Problem 1 Simulation Circuit}
\end{figure}

\begin{figure}[h]
    \centering
    \includegraphics[width = .6\linewidth]{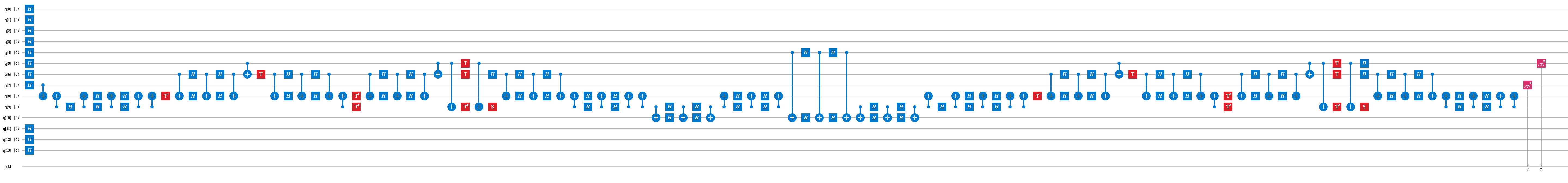}
    \caption{14-qubit Example Problem 1 Circuit on IBMQX16}
\end{figure}

\begin{figure}
    \centering
    \includegraphics[width = \linewidth]{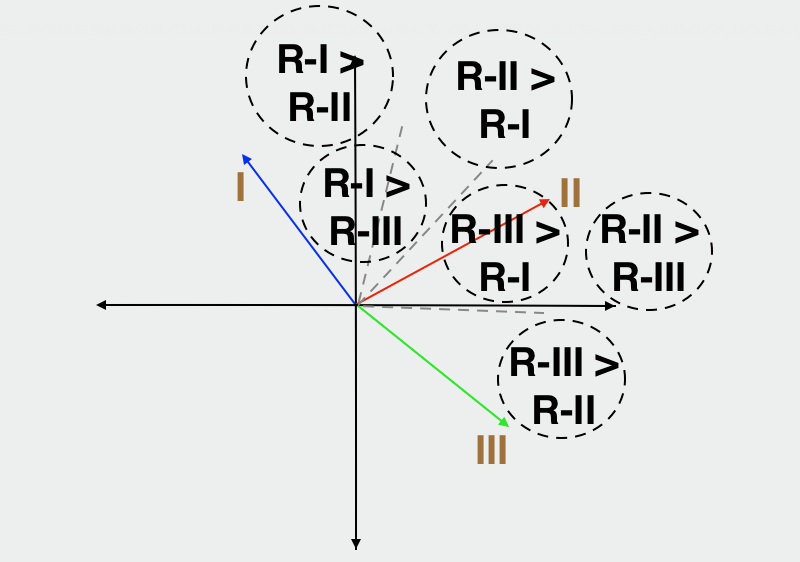}
    \caption{Preferred decision subregions for $> 2$ classes. \\
    The class vectors are shown in red, blue and green and indicated in roman numerals, whereas the grey dotted lines are the respective bisectors of each pair of class vectors. The dotted circles label the preferred subregions.}
\end{figure}
 
 \end{document}